# Spin-orbit torque engineering via oxygen manipulation

Xuepeng Qiu[1], Kulothungasagaran Narayanapillai[1], Yang Wu[1], Praveen Deorani[1], Dong-Hyuk Yang[2], Woo-Suk Noh[2], Jae-Hoon Park[2,3], Kyung-Jin Lee[4,5], Hyun-Woo Lee[6], and Hyunsoo Yang[1,*]

[1]Department of Electrical and Computer Engineering, and NUSNNI-Nanocore, National University of Singapore, 117576, Singapore

[2]c_CCMR and Department of Physics, Pohang University of Science and Technology, Pohang 790-784, Korea

[3]Division of Advanced Materials Science, Pohang University of Science and Technology, Pohang 790-784, Korea

[4]Department of Materials Science and Engineering, Korea University, Seoul 136-701, Korea

[5]KU-KIST Graduate School of Converging Science and Technology, Korea University, Seoul 136-713, Korea

[6]PCTP and Department of Physics, Pohang University of Science and Technology, Pohang 790-784, Korea

*e-mail: eleyang@nus.edu.sg

**Spin transfer torques allow the electrical manipulation of the magnetization at room temperature, which is desirable in spintronic devices such as spin transfer torque memories.**

**When combined with spin-orbit coupling, they give rise to spin-orbit torques which are a more powerful tool for magnetization control and can enrich device functionalities. The engineering of spin-orbit torques, based mostly on the spin Hall effect, is being intensely pursued. Here we report that the oxidation of spin-orbit torque devices triggers a new mechanism of spin-orbit torque, which is about two times stronger than that based on the spin Hall effect. We thus introduce a way to engineer spin-orbit torques via oxygen manipulation. Combined with electrical gating of the oxygen level, our findings may also pave the way towards reconfigurable logic devices.**

Controlling the magnetization direction via interaction between spins and charges is crucial for spintronic memory and logic devices[1-4]. Magnetization switching using conventional current-induced spin transfer torque (STT) requires a spin polarizer in a spin valve structure[5,6]. Recently, the combination of STT with spin-orbit coupling has led to a new type of torque, namely spin orbit torque (SOT). In magnetic bilayers, where an ultrathin ferromagnetic (FM) layer is in contact with a heavy metal (HM) layer, SOT arises from an in-plane current and enables efficient manipulation of the magnetization[7-14], in particular, low power magnetization switching[7,8], fast domain wall motion[9,15,16], and tunable nano-oscillators[11,12]. The high stability, simplicity, and scalability of SOT make it attractive for next-generation spintronic devices[8,17]. To fully realize the envisioned merits, enhanced control and design of SOT are desired.

So far, most experimental studies[8,16,18-20] have attributed SOT to the bulk spin Hall effect in the HM. According to this interpretation, the sign and magnitude of SOT are determined essentially by the spin Hall angle in the HM, while the characteristics of FM[18,21,22] may only slightly affect the magnitude of SOT. Here we report experimental results whose interpretation requires additional elements beyond the bulk spin Hall effect. We examine the oxidation effect on SOT in Pt/CoFeB/MgO/SiO$_2$. The Pt layer, which is the source of the bulk spin Hall effect, is hardly affected by oxygen in our experiment, and the oxygen effect is limited to the CoFeB layer. We find that as the oxygen level in the CoFeB layer goes above a threshold level, SOT suddenly reverses its direction while its magnitude remains roughly invariant. The bulk spin Hall interpretation of SOT may explain the magnitude change but is unable to explain the sign reversal. This result thus implies that a new SOT mechanism is introduced by the oxidation. We estimate that the new mechanism in our sample can be two times stronger than the bulk spin Hall mechanism, resulting in greater tunability of SOT in addition to the bulk spin Hall effect.

**Capping layer thickness dependence of SOT**

The oxygen level can be controlled by the thickness of the SiO$_2$ capping layer in our layer structure. The sputter-deposited film structure of Pt/CoFeB/MgO/SiO$_2$ is shown in Fig. 1a, in which the thickness $t$ of the SiO$_2$ layer is varied from 0 to 4 nm. For small $t$, oxygen can easily diffuse through both the SiO$_2$ and MgO layers, and reach the CoFeB layer. A scanning electron microscope (SEM) image of the patterned Hall bar is shown in Fig. 1b. We find that $t$ variation alters the SOT considerably. Figure 1c shows the anomalous Hall resistance ($R_H$) of the device as a function of the in-plane current ($I$) applied to the device. In addition to the current, a small constant magnetic field of 40 mT is applied along the positive current direction to break the symmetry[7,8] of the device and allow for selective magnetization switching by the in-plane current. Since $R_H$ probes the average $z$-component of the CoFeB magnetization, the hysteretic switching of $R_H$ confirms that the current-induced SOT indeed switches the magnetization. The arrows represent the current sweep direction. Interestingly, the resulting switching sequence is clockwise for $t > 1.5$ nm, but counterclockwise for $t \leq 1.5$ nm. Only the switching sequence for large $t$ (i.e., low oxygen level) is consistent with the previously reported switching sequence for Pt/Co/AlOx[7,23] and Pt/CoFe/MgO[16]. Thus, the switching sequence for small $t$ is abnormal.

The current-induced Oersted field cannot explain the sequence reversal. The switching sequence reversal is not due to the sign reversal of the relation between $R_H$ and the $z$-component of the magnetization either, since the purely magnetic-field driven magnetization switching (see Supplementary Figs. 2 and 6) does not exhibit the switching sequence reversal with $t$. The inset of Fig. 1e shows $I_S$ versus $t$, in which $I_S$ is defined as the current at which $R_H$ changes from a positive to a negative value (note the sign change of $I_S$ around $t = 1.5$ nm). Interestingly, this threshold thickness of 1.5 nm matches the native oxide thickness of Si for passivation. The main panel in Fig. 1e shows the ratio between the anisotropy field $H_{an}$ and $I_S$ as a function of $t$, which provides a rough magnitude estimation of SOT[17,23]. Note the abrupt sign reversal of the ratio while its magnitude remains roughly the same before and after the sign reversal. This implies that a new mechanism of SOT is suddenly introduced by the oxidation, generating SOT that is two times stronger and of opposite sign. The abrupt and full sign reversal differs qualitatively from continuous and marginal sign reversal in a previous study[21].

For independent confirmation of the SOT sign reversal, we perform lock-in measurements of SOT[13,21,24-26]. We apply a small amplitude sinusoidal ac current with a frequency of 13.7 Hz to exert periodic SOT on the magnetization, so that the induced magnetization oscillation around the equilibrium direction generates the second harmonic signal $V_{2\omega}$. Depending on the measurement geometry, $V_{2\omega}$ measures[21,25,26] the damping-like or field-like component of SOT, which are two mutually orthogonal vector components of SOT. As current induced magnetization switching is driven mainly by the damping-like SOT[7,8,23], we present here the results for the damping-like SOT only, which can be probed by applying an external dc magnetic field $H$ along the current direction (tilted 4 degree away from the film plane) to tilt the equilibrium magnetization direction accordingly. The results for the field-like SOT are presented in Supplementary Fig. 4. The magnetization switching characteristics have been measured from the first harmonic signals, and asymmetric $V_{2\omega}$ loops have been observed as shown in Fig. 1d. For $t = 0$ and 1.2 nm, there is a dip in $V_{2\omega}$ at a positive field and a peak at a negative field, while the opposite behaviour is observed for $t = 1.8$ and 3 nm. Opposite polarities in $V_{2\omega}$ prove that the damping-like SOT is pointing in opposite directions for small and large $t$, confirming the conclusion drawn from the switching sequence in Fig. 1c. For $t = 1.5$ nm, the $V_{2\omega}$ signal for positive field contains both a peak and a dip, which may indicate the coexistence of regions with opposite damping-like SOT directions. The extracted strengths of the damping-like effective field ($H_L$) are summarized in Fig. 1f for various $t$, indicating a sudden sign reversal of the damping-like SOT in agreement with Fig. 1e.

**Characterisation of oxidation**

In order to verify the oxidation for small $t$, we have carried out various measurements. Figure 2a shows the oxygen depth profiles obtained by secondary ion mass spectroscopy (SIMS) for devices with $t = 0$ and 2 nm. The depth profile for $t = 0$ nm shows a significantly enhanced oxygen level in the CoFeB layer compared to that of $t = 2$ nm, confirming the oxidation for small $t$. On the other hand, the two depth profiles are almost identical in the Pt layer, indicating no oxidation of the Pt layer even for small $t$. This is natural since Pt has excellent resistance to oxidation, which is supported also by the essentially indistinguishable Pt 4$f$ x-ray photoelectron spectroscopy (XPS) spectra in Fig. 2b for the $t = 0$ and 2 nm samples. We also use the x-ray absorption spectroscopy (XAS) to probe the electronic structures of Fe and Co. The Fe and Co $L_{2,3}$–edge (2$p$ → 3$d$) XAS spectra[27] in Fig. 2c and in Supplementary Fig. 10a exhibit spectral features[28] quite similar to those of α-Fe$_2$O$_3$ and CoO,[29,30] indicating (Fe,Co)–oxide formation with Fe$^{3+}$ and Co$^{2+}$, respectively. The XAS data show that the fraction of the oxidized atoms increases as the thickness $t$ decreases (Figs. 2f and Supplementary Fig. 11). Since Fe$^{3+}$ and Co$^{2+}$ ions do not have net magnetic moments, the saturation magnetization is expected to decrease with decreasing $t$, which is indeed the case as confirmed (Fig. 2e) by vibrating sample magnetometry (VSM). The magnetic properties of Fe and Co can also be probed by the x-ray magnetic circular dichroism (XMCD). Both the Fe (Fig. 2d) and Co (Supplementary Fig. 10b) $L_{2,3}$–edge XMCD signals become weaker with decreasing $t$, which is consistent with the VSM measurements, since the XMCD signals arise from the ferromagnetic atoms that remain un-ionized. The orbital magnetic moments of the ferromagnetic atoms can be evaluated from the XMCD sum rule.[27] For the ferromagnetic Fe atoms, the ratio between the orbital to the spin magnetic moments is about 0.06 (Fig. 2f), which is about 40% larger than the bulk value 0.043 and comparable to the value for epitaxial Fe film at the two-dimensional percolation threshold[27,31]. Interestingly, as $t$ decreases, the ratio increases even further to 0.065, indicating an enhancement of the orbital moment of the FM at the interface with the oxide[32]. Considering that orbital moment enhancement typically occurs when ferromagnetic atoms are in an environment with broken symmetry[33], this suggests that the ferromagnetic atoms are subject to more strongly broken symmetry as $t$ decreases.

**Sign reversal of SOT by *in-situ* oxidation**

All these measurements support the oxidation of the CoFeB layer for small $t$. However, we cannot yet rule out the possibility that the oxidation is merely correlated with instead of the cause of the SOT sign reversal. In order to verify that the oxidation of the FM is the key parameter for the sign reversal, we examine a sample with an oxidized CoFeB layer but with large $t$ (3 nm). To prepare such a sample, we intentionally oxidize the CoFeB layer with $O_2$ gas during its deposition before depositing the capping layers including 3 nm of $SiO_2$ as shown in Fig. 3a. In this case, even for $t = 3$ nm, we observe an abnormal anticlockwise switching loop as shown in Fig. 3b. Hence what is important for the SOT sign reversal is the FM layer oxidation itself rather than the value of $t$. Small $t$ is merely a method to induce the oxidation. This result rules out other $t$-dependent changes such as strain from being key parameters of the SOT sign reversal.

**SOT beyond the spin Hall effect**

We now discuss the connection between the FM layer oxidation and the SOT sign reversal. The bulk spin Hall effect[8] in the HM layer is an important mechanism of SOT. According to the spin Hall interpretation of SOT[16,23], the sign of the damping-like SOT is determined by the bulk spin Hall angle of the HM layer. The FM affects the damping-like SOT through the real part of the spin mixing conductance[34], which is always positive and cannot change its sign since its being negative implies a negative charge conductance.[35] The spin Hall interpretation is thus inadequate to explain the oxygen-induced sign reversal of the damping-like SOT, since the HM layer is not affected by oxidation (Fig. 2a and b). Furthermore, we have verified that the SOT change caused by the oxidation is essentially independent of the Pt thickness (Supplementary Fig. 5). Hence our data necessitate a new source of SOT, other than the bulk spin Hall effect of the HM layer. One can think of two possibilities; one is the oxidized FM layer itself being a SOT source, and the other is the top or bottom interfaces of the oxidized FM layer being a SOT source.

To examine the first possibility, we change the thickness $d_{CFB}$ of the CoFeB layer for fixed $t = 0$ nm. Up to $d_{CFB} = 2$ nm, the perpendicular magnetic anisotropy is well maintained, and the saturation magnetization ($M_S$) values (Fig. 4a) are almost constant and do not increase with $d_{CFB}$, implying an almost $d_{CFB}$-independent oxidation level. The change of $d_{CFB}$ has a negligible effect on the current density, since the resistivity of CoFeB is much greater than that of Pt[22]. While the abnormal anti-clockwise switching sequence is maintained (Fig. 4b) with changing $d_{CFB}$, $H_{an}/I_S$ and $H_L$ change almost linearly with $1/d_{CFB}$ (Fig. 4c). That is, $H_L \times d_{CFB}$, which is proportional to the total torque acting on the CoFeB layer, does not increase with $d_{CFB}$. This implies that the bulk part of the oxidized FM layer is not an SOT source. Hence, one can exclude the first possibility.

To examine the second possibility, we eliminate the MgO layer from the device stack structure. The switching sequence reversal from the normal clockwise to abnormal anti-clockwise direction is still observed, when $t$ changes from 4 to 1.2 nm as shown in Fig. 5a. This shows that the interface between the oxidized FM layer and the MgO layer is not the new SOT source. Next we eliminate the Pt layer instead. The current-induced switching itself (Fig. 5b) is not observed nor is the 2$^{nd}$ harmonic signal (Supplementary Fig. 13). This leads us to conclude that the interface between the oxidized CoFeB layer and the Pt layer is the new SOT source. We further extend our experiments to devices in which the FM material CoFeB is replaced with Co. As shown in Fig. 5c, the current-induced switching sequence shows normal clockwise behaviour for $t = 3$ nm, but reverses to abnormal anti-clockwise behaviour for $t = 0$ nm, which is similar to the results from devices with CoFeB as the FM layer. Hence the observed sign reversal phenomenon is not restricted to a specific FM material, but can be universal.

The most plausible mechanism consistent with our experimental data is then the interfacial spin-orbit coupling[34,36-40], some signatures of which have been reported in earlier experiments by varying the degree of oxidation or changing the thickness of the Ta underlayer[7,21].

If its contribution to the damping-like SOT is of opposite sign to the spin Hall contribution and becomes larger with oxidation than the spin Hall contribution, the competition between these two contributions can explain the sign reversal of the damping-like SOT upon oxidation. For this, the oxidation should enhance either (i) the interfacial spin-orbit coupling strength or (ii) the efficiency to generate the damping-like torque for a given interfacial spin-orbit coupling strength.

There is a well-known example of (i). The interfacial spin-orbit coupling at the magnetic Gd(0001) surface[41] becomes three times stronger upon oxidation, and interestingly reverses its sign. The enhanced strength is attributed to the enhanced internal electric field at the surface. An additional mechanism of (i) may arise from the atomic orbital degree of freedom. When atomic orbitals with angular momentum $\vec{L}$ are linearly superposed to make a Bloch state with crystal momentum $\vec{K}$, the quantum interference between orbitals of neighbouring atoms generates an electric dipole moment[42] towards the direction $\vec{L}\times\vec{K}$, which couples with the internal electric field $\vec{E}$ at the surface to generate a Coulomb energy proportional to $-\vec{E}\cdot(\vec{L}\times\vec{K})$. When $\vec{E}$ is sufficiently strong and the orbital quenching is weak, this energy tends to align[43] $\vec{L}$ along the direction $\vec{K}\times\vec{E}$. Such $\vec{L}$ couples with spin $\vec{S}$ through the atomic spin-orbit coupling $\alpha_{SO}\vec{L}\cdot\vec{S}$ at the surface, where the coupling constant $\alpha_{SO}$ is large due to the hybridization between Pt $5d$ orbitals and ferromagnetic $3d$ orbitals[44,45]. Subsequently, the strong atomic spin-orbit coupling $\alpha_{SO}\vec{L}\cdot\vec{S}$ is converted to the strong interfacial spin-orbit coupling $\alpha_{SO}\vec{K}\times\vec{E}\cdot\vec{S}$. It has been suggested[42,44] that this orbital-based mechanism may enhance the Rashba-type interfacial spin-orbit coupling significantly. A recent first principles calculation[45] for a Pt/Co bilayer confirms that a strong interfacial spin-orbit coupling can indeed arise at the HM/FM interface near the Fermi energy.

Regarding (ii), we are not aware of any theoretical mechanism that predicts efficiency enhancement by oxidation. We remark, however, that the damping-like SOT caused by interfacial spin-orbit coupling has been significantly underestimated in earlier theories[34,36,37,39]. It was later pointed out[38,46] that due to the Berry phase effect, the efficiency of the interfacial spin-orbit coupling mechanism is actually much higher and comparable to that of the spin Hall mechanism. The Berry phase effect has been confirmed[46] for a bulk inversion symmetry broken material (Ga,Mn)As, but not yet for structural inversion symmetry broken interfaces. Previous observations[11,15,16,18-20,23] of the damping-like SOT were attributed to the bulk spin Hall mechanism.

Next we discuss the abruptness of the SOT sign reversal (Figs. 1e & 1f) despite the rather gradual changes of oxidation level (Figs. 2f & Supplementary Fig. 11). Concomitant with the sudden SOT sign reversal, the coercivity (Supplementary Fig. 2) and the temperature dependence of $R_H$ (Supplementary Fig. 14) also change suddenly. We suspect that such sudden changes may be manifestations of SOT instability. A possible origin of the SOT instability is the competition between the orbital ordering $\vec{L} \propto \vec{K} \times \vec{E}$ and the orbital quenching common in transition metals. Although its origin is still unclear, the abrupt SOT reversal is of considerable value for device applications. When the oxidation level is near the threshold value, a tiny change of the oxidation level by electric gating[47,48] can induce a large change of SOT. This takes the SOT engineering to a whole new level and may even pave the way towards reconfigurable logic devices.

Our results may also be relevant to recent SOT experiments using Ta[8,16,21,49] which is more susceptible to oxidation compared to Pt. Our results indicate that even for the exact same layer structure, very different SOTs can be obtained depending on the detailed device preparation procedure, which may affect the oxygen content in the sample. Furthermore, we hope our work initiates efforts to bridge the gap between metal spintronics and oxide electronics to combine the merits of the both fields[50].

**Methods**

The stacked films were deposited on thermally oxidized Si substrates by magnetron sputtering with a base pressure $< 2 \times 10^{-9}$ Torr at room temperature. The structure of the $t$ series films is substrate/MgO (2)/Pt (2)/Co$_{60}$Fe$_{20}$B$_{20}$ (0.8)/MgO (2)/SiO$_2$ ($t$) with $t = 0 \sim 4$, and that of the $d_{CFB}$ series films is substrate/MgO (2)/Pt (2)/Co$_{60}$Fe$_{20}$B$_{20}$ ($d_{CFB}$)/MgO (2)/SiO$_2$ ($t = 0$ and 3) with $d_{CFB} = 0.8 \sim 2$ (numbers are nominal thicknesses in nanometers). The bottom MgO layer is used to promote perpendicular anisotropy. The other film structures are schematically shown in Figs. 3 and 5. After deposition, except for the oxygen doped CoFeB sample in Fig. 3, all the other films were post-annealed at 300 °C for 1 hour in a vacuum to obtain perpendicular anisotropy. The multilayers were coated with a ma-N 2401 negative e-beam resist and patterned into 600 nm width Hall bars by electron beam lithography and Ar ion milling as shown in Fig. 1b. PG remover and acetone were used to lift-off the e-beam resist. Contact pads were defined by photolithography followed by the deposition of Ta (5 nm)/Cu (150 nm)/Ru (5 nm) which are connected to the Hall bars. Before the deposition of the contact pads, Ar ion milling was used to remove the SiO$_2$ and part of the MgO layer, in order to make low-resistance electrical contacts. All the devices for each batch were processed at the same time to ensure the same fabrication conditions, which was important for this study. Devices were wire-bonded to the sample holder and installed in a physical property measurement system (Quantum Design) for the transport studies.

We performed the measurements of current induced switching and the ac harmonic anomalous Hall voltage loops for the $t$ and $d_{CFB}$ series devices at 200 K for the data set in Figs. 1 and 4, at which temperature all the devices retain desirable perpendicular anisotropy (Supplementary Figs. 1-3). The current induced switching was measured using a combination of Keithley 6221 and 2182A. A pulsed dc current of a duration of 50 μs was applied to the nanowires and the Hall voltage was measured simultaneously. An interval of 0.1 s was used for the pulsed dc current to eliminate the accumulated Joule heating effect.

The x-ray absorption spectroscopy (XAS) and x-ray magnetic circular dichroism (XMCD) measurements were performed at the 2A beamline in the Pohang Light Source. All the spectra were measured at 200 K in the total electron yield mode, and the energy resolution was set to be ~ 300 meV. The reference XAS spectra were obtained from $\alpha$-$Fe_2O_3$ and CoO bulk crystals at room temperature. An electromagnet with a 0.4 T magnetic field was used for the magnetization along the normal direction of the sample for the XMCD measurements at normal incidence of more than 95% circularly polarized light. The spin and orbital magnetic moment ratio was estimated by using the sum rule[27], and the metal to the oxide ratio was extracted from the XAS spectra by using the reference spectra of metallic Fe (Co) and $\alpha$-$Fe_2O_3$ (CoO).

# References

1. Wolf, S. A. *et al.* Spintronics: A spin-based electronics vision for the future. *Science* **294**, 1488-1495 (2001).
2. Brataas, A., Kent, A. D. & Ohno, H. Current-induced torques in magnetic materials. *Nat. Mater.* **11**, 372-381 (2012).
3. Zutic, I., Fabian, J. & Das Sarma, S. Spintronics: Fundamentals and applications. *Rev. Mod. Phys.* **76**, 323-410 (2004).
4. Awschalom, D. D. & Flatte, M. E. Challenges for semiconductor spintronics. *Nat. Phys.* **3**, 153-159 (2007).
5. Katine, J. A., Albert, F. J., Buhrman, R. A., Myers, E. B. & Ralph, D. C. Current-Driven Magnetization Reversal and Spin-Wave Excitations in Co /Cu /Co Pillars. *Phys. Rev. Lett.* **84**, 3149 (2000).
6. Ralph, D. C. & Stiles, M. D. Spin transfer torques. *J. Magn. Magn. Mater.* **320**, 1190-1216 (2008).
7. Miron, I. M. *et al.* Perpendicular switching of a single ferromagnetic layer induced by in-plane current injection. *Nature* **476**, 189-193 (2011).
8. Liu, L. Q. *et al.* Spin-Torque Switching with the Giant Spin Hall Effect of Tantalum. *Science* **336**, 555-558 (2012).
9. Miron, I. M. *et al.* Fast current-induced domain-wall motion controlled by the Rashba effect. *Nat. Mater.* **10**, 419-423 (2011).
10. Miron, I. M. *et al.* Current-driven spin torque induced by the Rashba effect in a ferromagnetic metal layer. *Nat. Mater.* **9**, 230-234 (2010).
11. Liu, L. Q., Pai, C. F., Ralph, D. C. & Buhrman, R. A. Magnetic Oscillations Driven by the Spin Hall Effect in 3-Terminal Magnetic Tunnel Junction Devices. *Phys. Rev. Lett.* **109**, 186602 (2012).
12. Demidov, V. E. *et al.* Magnetic nano-oscillator driven by pure spin current. *Nat. Mater.* **11**, 1028-1031 (2012).
13. Jamali, M. *et al.* Spin-Orbit Torques in Co/Pd Multilayer Nanowires. *Phys. Rev. Lett.* **111**, 246602 (2013).
14. Suzuki, T. *et al.* Current-induced effective field in perpendicularly magnetized Ta/CoFeB/MgO wire. *Appl. Phys. Lett.* **98**, 142505 (2011).
15. Ryu, K.-S., Thomas, L., Yang, S.-H. & Parkin, S. Chiral spin torque at magnetic domain walls. *Nat. Nanotechnol.* **8**, 527-533 (2013).
16. Emori, S., Bauer, U., Ahn, S. M., Martinez, E. & Beach, G. S. Current-driven dynamics of chiral ferromagnetic domain walls. *Nat. Mater.* **12**, 611-616 (2013).
17. Lee, K.-S., Lee, S.-W., Min, B.-C. & Lee, K.-J. Threshold current for switching of a perpendicular magnetic layer induced by spin Hall effect. *Appl. Phys. Lett.* **102**, 112410-112415 (2013).
18. Fan, X. *et al.* Observation of the nonlocal spin-orbital effective field. *Nat. Commun.* **4**, 1799 (2013).
19. Pai, C.-F. *et al.* Spin transfer torque devices utilizing the giant spin Hall effect of tungsten. *Appl. Phys. Lett.* **101**, 122404 (2012).
20. Haazen, P. P. J. *et al.* Domain wall depinning governed by the spin Hall effect. *Nat. Mater.* **12**, 299-303 (2013).
21. Kim, J. *et al.* Layer thickness dependence of the current-induced effective field vector in Ta|CoFeB|MgO. *Nat. Mater.* **12**, 240-245 (2013).
22. Fan, X. *et al.* Quantifying interface and bulk contributions to spin-orbit torque in magnetic bilayers. *Nat. Commun.* **5**, 3042 (2014).

23  Liu, L. Q., Lee, O. J., Gudmundsen, T. J., Ralph, D. C. & Buhrman, R. A. Current-Induced Switching of Perpendicularly Magnetized Magnetic Layers Using Spin Torque from the Spin Hall Effect. *Phys. Rev. Lett.* **109**, 096602 (2012).
24  Pi, U. H. *et al.* Tilting of the spin orientation induced by Rashba effect in ferromagnetic metal layer. *Appl. Phys. Lett.* **97**, 162507 (2010).
25  Garello, K. *et al.* Symmetry and magnitude of spin-orbit torques in ferromagnetic heterostructures. *Nat. Nanotechnol.* **8**, 587-591 (2013).
26  Qiu, X. *et al.* Angular and temperature dependence of current induced spin-orbit effective fields in Ta/CoFeB/MgO nanowires. *Sci. Rep.* **4**, 4491 (2014).
27  Chen, C. T. *et al.* Experimental Confirmation of the X-Ray Magnetic Circular Dichroism Sum Rules for Iron and Cobalt. *Phys. Rev. Lett.* **75**, 152-155 (1995).
28  de Groot, F. M. F., Fuggle, J. C., Thole, B. T. & Sawatzky, G. A. 2p x-ray absorption of 3d transition-metal compounds: An atomic multiplet description including the crystal field. *Phys. Rev. B* **42**, 5459-5468 (1990).
29  Kim, J. Y. *et al.* Ferromagnetism Induced by Clustered Co in Co-Doped Anatase TiO2 Thin Films. *Phys. Rev. Lett.* **90**, 017401 (2003).
30  Park, S. *et al.* Strain control of Morin temperature in epitaxial α-Fe2O3(0001) film. *Europhys. Lett* **103**, 27007 (2013).
31  Ohresser, P., Ghiringhelli, G., Tjernberg, O., Brookes, N. B. & Finazzi, M. Magnetism of nanostructures studied by x-ray magnetic circular dichroism:Fe on Cu(111). *Phys. Rev. B* **62**, 5803-5809 (2000).
32  Nistor, C. *et al.* Orbital moment anisotropy of Pt/Co/AlOx heterostructures with strong Rashba interaction. *Phys. Rev. B* **84**, 054464 (2011).
33  Wu, Y., Stöhr, J., Hermsmeier, B. D., Samant, M. G. & Weller, D. Enhanced orbital magnetic moment on Co atoms in Co/Pd multilayers: A magnetic circular x-ray dichroism study. *Phys. Rev. Lett.* **69**, 2307-2310 (1992).
34  Haney, P. M., Lee, H.-W., Lee, K.-J., Manchon, A. & Stiles, M. D. Current induced torques and interfacial spin-orbit coupling: Semiclassical modeling. *Phys. Rev. B* **87**, 174411 (2013).
35  Brataas, A., Nazarov, Y. V. & Bauer, G. E. W. Finite-Element Theory of Transport in Ferromagnet–Normal Metal Systems. *Phys. Rev. Lett.* **84**, 2481-2484 (2000).
36  Pesin, D. A. & MacDonald, A. H. Quantum kinetic theory of current-induced torques in Rashba ferromagnets. *Phys. Rev. B* **86**, 014416 (2012).
37  Wang, X. H. & Manchon, A. Diffusive Spin Dynamics in Ferromagnetic Thin Films with a Rashba Interaction. *Phys. Rev. Lett.* **108**, 117201 (2012).
38  van der Bijl, E. & Duine, R. A. Current-induced torques in textured Rashba ferromagnets. *Phys. Rev. B* **86**, 094406 (2012).
39  Kim, K.-W., Seo, S.-M., Ryu, J., Lee, K.-J. & Lee, H.-W. Magnetization dynamics induced by in-plane currents in ultrathin magnetic nanostructures with Rashba spin-orbit coupling. *Phys. Rev. B* **85**, 180404 (2012).
40  Haney, P. M. & Stiles, M. D. Current-Induced Torques in the Presence of Spin-Orbit Coupling. *Phys. Rev. Lett.* **105**, 126602 (2010).
41  Krupin, O. *et al.* Rashba effect at magnetic metal surfaces. *Phys. Rev. B* **71**, 201403 (2005).
42  Park, S. R., Kim, C. H., Yu, J., Han, J. H. & Kim, C. Orbital-angular-momentum based origin of Rashba-type surface band splitting. *Phys. Rev. Lett.* **107**, 156803 (2011).
43  Park, S. R. *et al.* Chiral orbital-angular momentum in the surface states of Bi2Se3. *Phys. Rev. Lett.* **108**, 046805 (2012).

44	Park, J.-H., Kim, C. H., Lee, H.-W. & Han, J. H. Orbital chirality and Rashba interaction in magnetic bands. *Phys. Rev. B* **87**, 041301 (2013).
45	Haney, P. M., Lee, H.-W., Lee, K.-J., Manchon, A. & Stiles, M. D. Current-induced torques and interfacial spin-orbit coupling. *Phys. Rev. B* **88**, 214417 (2013).
46	KurebayashiH *et al.* An antidamping spin-orbit torque originating from the Berry curvature. *Nat. Nanotechnol.* **9**, 211-217 (2014).
47	Jeong, J. *et al.* Suppression of Metal-Insulator Transition in VO2 by Electric Field–Induced Oxygen Vacancy Formation. *Science* **339**, 1402-1405 (2013).
48	Bauer, U., Emori, S. & Beach, G. S. Voltage-controlled domain wall traps in ferromagnetic nanowires. *Nat. Nanotechnol.* **8**, 411-416 (2013).
49	Yu, G. *et al.* Switching of perpendicular magnetization by spin-orbit torques in the absence of external magnetic fields. *Nat. Nanotechnol.* **9**, 548-554 (2014).
50	Narayanapillai, K. *et al.* Current-driven spin orbit field in LaAlO3/SrTiO3 heterostructures. *Appl. Phys. Lett.* **105**, 162405 (2014).

**Acknowledgments**

This research is supported by the National Research Foundation, Prime Minister's Office, Singapore under its Competitive Research Programme (CRP Award No. NRF-CRP12-2013-01 and NRF-CRP4-2008-06). K.L. acknowledges financial support by the NRF (NRF-2013R1A2A2A01013188) and the MEST Pioneer Research Center Program (2011-0027905). H.W.L acknowledges financial support by the NRF (NRF-2011-0030046 and NRF-2013R1A2A2A05006237) and MOTIE (10044723). D.Y, W.N, and J.P acknowledge financial support by NCRI Program (2009-0081576) and MPK Program (2011-0031558) through NRF funded by the MSIP. PAL is supported by POSTECH and MSIP.

**Author contributions**

X.Q. and H.Y. planned the study. X.Q. and K.N. fabricated devices. X.Q. measured transport properties. Y.W. helped characterization. D.Y, W.N, and J.P carried out x-ray measurements. All authors discussed the results. X.Q., K.L., H.L., and H.Y. wrote the manuscript. H.Y. supervised the project.

**Additional information**

Supplementary information accompanies this paper at www.nature.com/naturenanotechnology. Reprints and permission information is available online at http://npg.nature.com/reprintsandpermissions/. Correspondence and requests for materials should be addressed to H. Y.

**Competing financial interests**

The authors declare no competing financial interests.

**Fig. 1. Device structure and the effect of the SiO$_2$ capping layer thickness. a**, Film structure of the multilayers. **b**, Scanning electron micrograph of the device and electrical measurement schematic. Two electrodes for the current injection are labelled with I+ and I-. The other two electrodes for the Hall voltage measurements are labelled with V+ and V-. **c**, $R_H$ as a function of pulsed currents for different $t$ at 200 K. $t$ is indicated in each graph. A fixed 40 mT magnetic field is applied along the positive current direction. The width of pulsed currents is 50 µs, and $R_H$ is measured after a 16 µs delay. **d**, The first ($V_\omega$) and second ($V_{2\omega}$) harmonic anomalous Hall voltages as a function of magnetic field ($H$). $H$ is applied along the current direction with a tilt of 4° away from the film plane and $I_{ac}$ = 141.4 µA. **e**, Anisotropy field $H_{an}$ divided by switching current $I_S$ (defined as the switching current from a high to low $R_H$) as a function of $t$. The inset shows $I_S$. **f**, The extracted damping-like effective field ($H_L$) from the 2$^{nd}$ harmonic data at positive magnetic fields with different $t$. The planar Hall effect was taken into consideration for the extraction of $H_L$ using the method in ref. 26. The solid lines in **e** and **f** are the B-spline interpolations to the experimental data.

**Fig. 2. SIMS, XPS, XAS, VSM, and XMCD characterisation. a**, SIMS depth profiles of oxygen for the films without ($t$ = 0 nm, solid line) and with ($t$ = 2 nm, dashed line) the SiO$_2$ capping layer. Expected layers with the etching time are denoted at the bottom of the graph. **b**, Pt 4$f$ XPS spectra for the films without ($t$ = 0 nm, solid line) and with ($t$ = 2 nm, dashed line) the SiO$_2$ capping. **c**, Fe $L_{2,3}$–edge XAS with various $t$ at 200 K. Peaks of Fe and Fe$_2$O$_3$ are indicated by arrows. **d**, Fe $L_{2,3}$–edge XMCD with various $t$ at 200 K. **e**, The saturation magnetization data by VSM at 200 K. The solid line is a B-spline interpolation to the experimental data. **f**, The orbital-to-spin moment ratio and the estimated $Fe^{3+}$–oxide site ratio as a function of $t$. The error bars correspond to uncertainties in the estimation of the ratios from the experimental data.

**Fig. 3. The CoFeB oxidation effect. a**, Device structure where the CoFeB layer is intentionally oxidized with reactive sputtering (20 sccm Ar + 10 sccm $O_2$) for the first 14 seconds, then deposited using only Ar for the rest of the CoFeB layer. The whole 1.6 nm CoFeB layer requires 2 minutes for deposition. **b,** Current-induced switching measured by $R_H$ as a function of applied current at 300 K. A magnetic field of $H = 4$ mT is applied along the positive current direction.

**Fig. 4. The CoFeB thickness effect. a**, The saturation magnetization data at 200 K for various thicknesses $d_{CFB}$ of CoFeB for $t = 0$ nm (diamond symbols). Data for $t = 3$ nm (square symbols) are presented as a reference. The solid lines are the B-spline interpolations to the experimental data. **b**, Current-induced switching for the various $d_{CFB}$ measured by $R_H$ as a function of applied current at 200 K. A magnetic field of $H = 40$ mT is applied along the positive current direction. **c**, $H_{an}/I_S$ and $H_L$ versus $1/d_{CFB}$. The solid lines show the linear fitting for the data.

**Fig. 5. Material and structural dependence of the sign reversal. a**, The MgO layer between the CoFeB and $SiO_2$ layers is eliminated from the stack. **b**, The Pt layer is eliminated. **c**, Co is utilized in the multilayer stack. The experimental setup is the same as that of Fig. 1c. The measurements were performed at 200 K for **a** and **c**, and at 300 K for **b**.

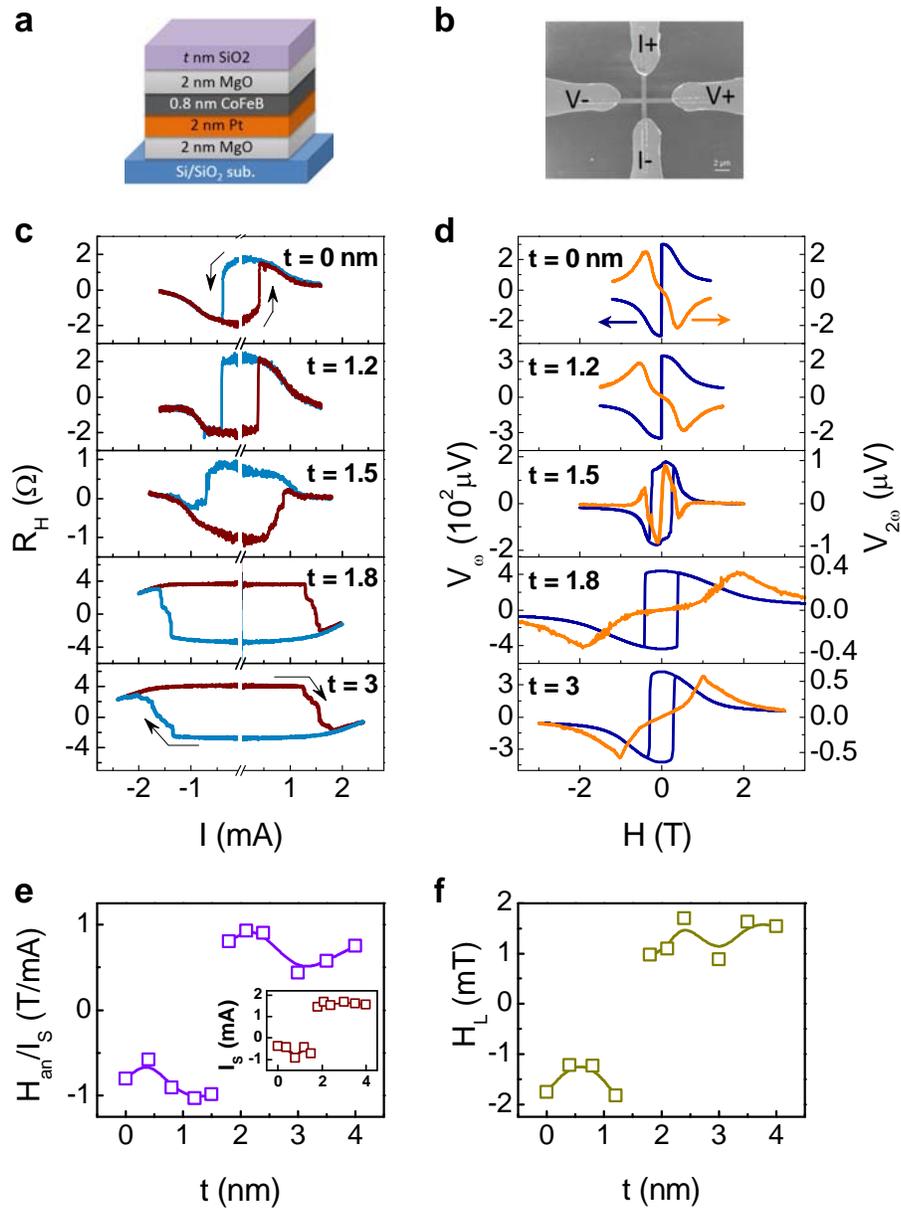

Fig. 1

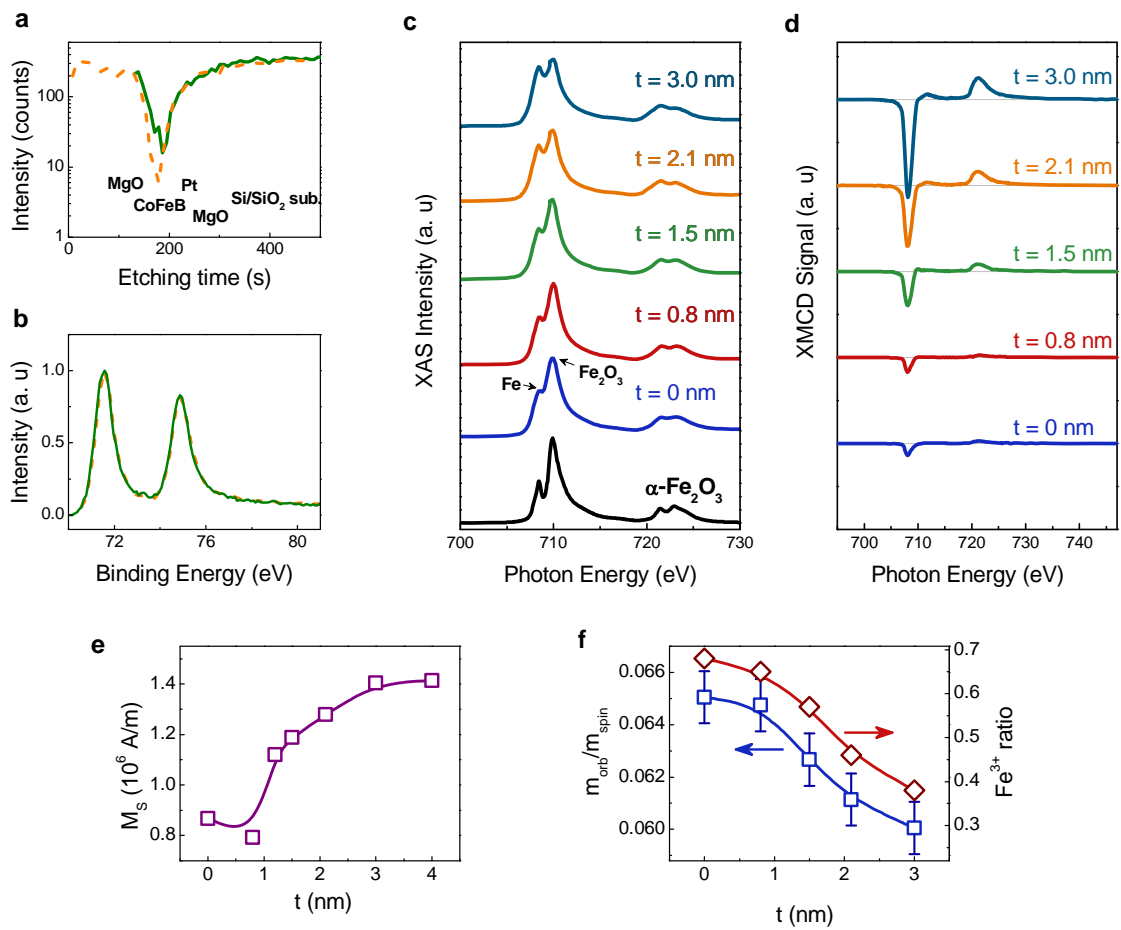

Fig. 2

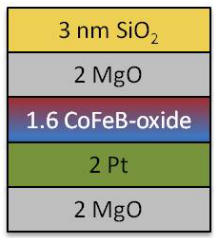
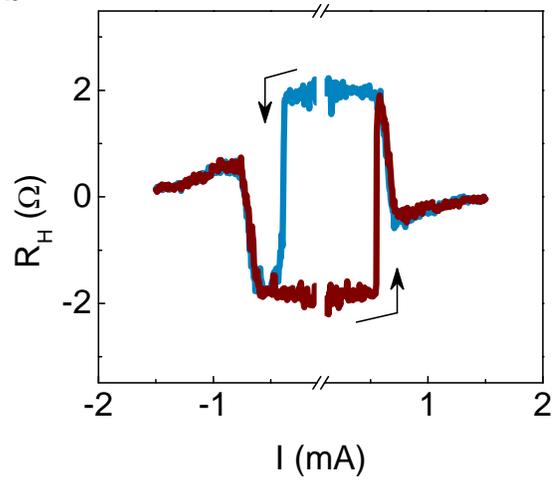

Fig. 3

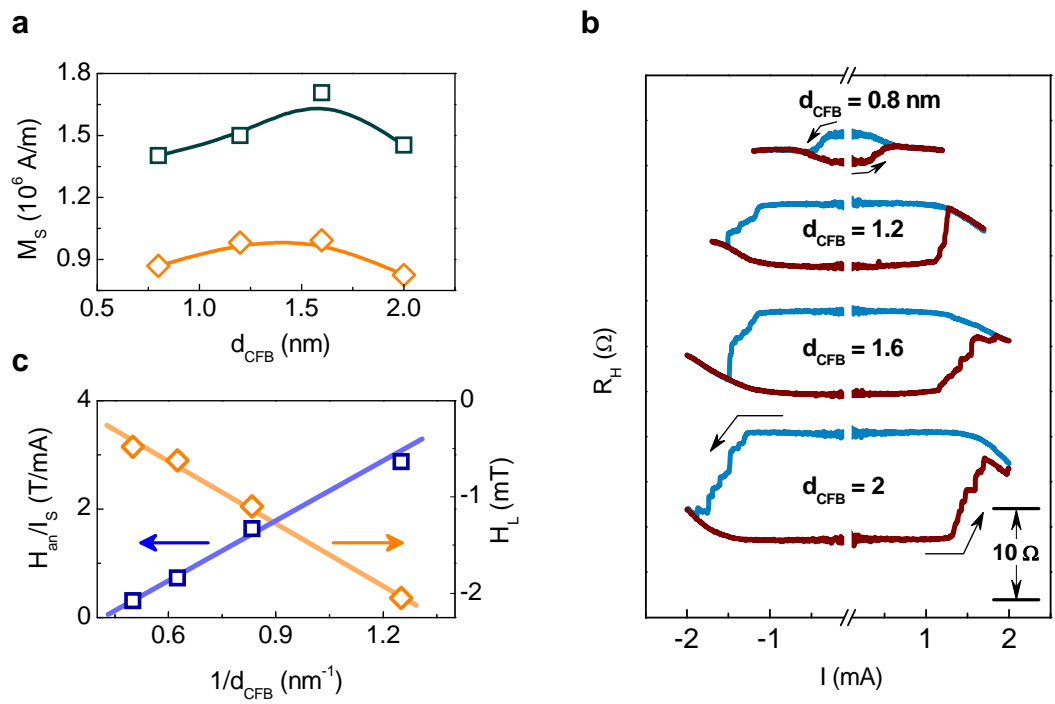

Fig. 4

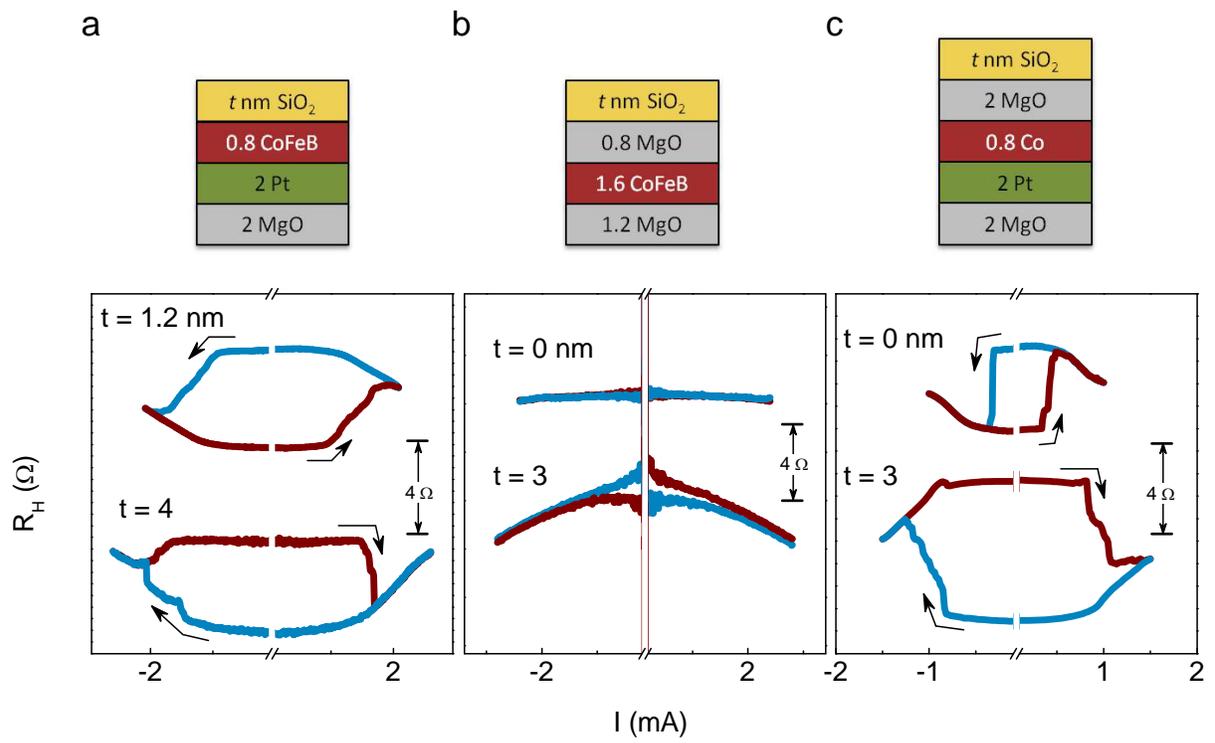

Fig. 5